\documentclass[aps,prl,reprint,superscriptaddress]{revtex4-1}
\usepackage{lipsum}

\usepackage{epsfig}
\usepackage{latexsym}
\usepackage{epic}
\usepackage{eepic}
\usepackage{psfrag}
\usepackage{rotating}
\usepackage{multirow,booktabs}
\usepackage{color}
\usepackage{cprotect} 
\usepackage{miller} 
\usepackage{xspace} 

\usepackage[export]{adjustbox}
\usepackage{graphicx}
 
\usepackage{algorithm}
\usepackage[]{algorithmicx,algpseudocode}

\usepackage[fleqn, leqno]{amsmath}
   \usepackage{amsfonts}   
   \usepackage{amssymb}    
\usepackage{amsthm}
\usepackage{stmaryrd} 
\usepackage{import}
\usepackage{breqn} 

\usepackage{array}

\usepackage{ulem}

\newcommand{\done}[1][]{{\color{green} \ensuremath{ \surd}\ifx\relax#1\relax\else(#1)\fi}} 
\newcommand{\prob}[1][]{{\color{red} { \bf!}\ifx\relax#1\relax\else(#1)\fi}} 
\newcommand{\fail}[1][]{{\color{red} \ensuremath{\lightning}\ifx\relax#1\relax\else(#1)\fi}} 

\newcommand{\ddr}{\ensuremath{{\rm d}{\bf r} }\xspace} 
\newcommand{\rr}{\ensuremath{{\bf r} }\xspace}

\newcommand{\mm}{\ensuremath{{\bf m} }\xspace} 
 
\newcommand{\mmi}{\ensuremath{{ m_i} }\xspace} 

\newcommand{\BB}{\ensuremath{{\bf B} }\xspace} 
\newcommand{\BBext}{\ensuremath{{\bf B_{\rm ext}} }\xspace} 
\newcommand{\NBext}{\ensuremath{{B_{\rm ext}} }\xspace} 
\newcommand{\BBind}{\ensuremath{{\bf B_{\rm ind}} }\xspace}

\newcommand{\VA}{\ensuremath{{\bf A} }\xspace} 

\newcommand{\phii}[1][]{\ensuremath{\varphi}\xspace\ifx\relax#1\relax\else\ensuremath{\left(#1\right)}\xspace\fi}

\newcommand{\FF}[1][]{\ensuremath{\mathcal{F}}\xspace\ifx\relax#1\relax\else\ensuremath{\left[#1\right]}\xspace\fi}
\newcommand{\FFm}[1][]{\ensuremath{\mathcal{F_{\rm m}}}\xspace\ifx\relax#1\relax\else\ensuremath{\left[#1\right]}\xspace\fi}
\newcommand{\FFpfc}[1][]{\ensuremath{\mathcal{F_{\rm PFC}}}\xspace\ifx\relax#1\relax\else\ensuremath{\left[#1\right]}\xspace\fi}
\newcommand{\FFcoup}[1][]{\ensuremath{\mathcal{F_{\rm c}}}\xspace\ifx\relax#1\relax\else\ensuremath{\left[#1\right]}\xspace\fi}

\newcommand{\ff}[1][]{\ensuremath{f}\ifx\relax#1\relax\else\ensuremath{\left(#1\right)}\fi}

\newcommand{\ffpfc}[1][]{\ensuremath{\ff_{\rm PFC}}\ifx\relax#1\relax\else\ensuremath{\left(#1\right)}\fi}
\newcommand{\ffid}[1][]{\ensuremath{\ff_{\rm id}}\ifx\relax#1\relax\else\ensuremath{\left(#1\right)}\fi}
\newcommand{\ffex}[1][]{\ensuremath{\ff_{\rm ex}}\ifx\relax#1\relax\else\ensuremath{\left(#1\right)}\fi}
\newcommand{\ffm}[1][]{\ensuremath{\ff_{\rm m}}\ifx\relax#1\relax\else\ensuremath{\left(#1\right)}\fi}
\newcommand{\ffcoup}[1][]{\ensuremath{\ff_{\rm c}}\ifx\relax#1\relax\else\ensuremath{\left(#1\right)}\fi}

\newcommand{\NN}[1][]{\ensuremath{\mathcal{N}}\ifx\relax#1\relax\else\ensuremath{\left[{#1}\right]}\fi}

\newcommand{\LL}[1][]{\ensuremath{\mathcal{L}}\ifx\relax#1\relax\else\ensuremath{#1}\fi}

\newcommand{\np}[1][]{\ensuremath{#1}\ifx\relax#1\relax\else\ensuremath{^{n+1}}\fi}
\newcommand{\n}[1][]{\ensuremath{#1}\ifx\relax#1\relax\else\ensuremath{^{n}}\fi}

\newcommand \be {\begin{eqnarray}}
\newcommand \ee {\end{eqnarray}}

\usepackage{tikz}
\usetikzlibrary{fadings,shapes.arrows,shadows}

\usepackage{xparse}
\graphicspath{{logo/} {figure/} {video/}}

\tikzfading[name=arrowfading, top color=transparent!0, bottom color=transparent!95]
\tikzset{arrowfill/.style={#1,general shadow={fill=black, shadow yshift=-0.ex, path fading=arrowfading}}}
\tikzset{arrowstyle/.style n args={3}{draw=#2,arrowfill={#3}, single arrow,minimum height=#1, single arrow,
single arrow head extend=.3cm,}}

\NewDocumentCommand{\tikzfancyarrow}{O{2cm} O{gray!10} O{top color=gray!20, bottom color=gray} m}{
\tikz[baseline=-0.5ex]\node [arrowstyle={#1}{#2}{#3}] {#4};
}

\makeatletter
\let\cat@comma@active\@empty
\makeatother

\begin{document}


\title{Controlling grain boundaries by magnetic fields}


\author{R. Backofen}
\affiliation{Institute of Scientific Computing, Technische Universit\"at Dresden, 01062 Dresden, Germany}
\author{K.R. Elder}
\affiliation{Department of Physics, Oakland University, Rochester, Michigan 48309, USA
 }
\author{A. Voigt}
\affiliation{Institute of Scientific Computing, Technische Universit\"at Dresden, 01062 Dresden, Germany}
\affiliation{Dresden Center for Computational Materials Science (DCMS), 01062 Dresden, Germany}

\begin{abstract}
The ability to use external magnetic fields to influence the microstructure in
polycrystalline materials has potential applications in microstructural
engineering. To explore this potential and to understand the complex
interactions between electromagnetic fields and solid-state matter transport we
consider a phase-field-crystal (PFC) model. Together with efficient and
scalable numerical algorithms this allows the examination of the role that external
magnetic fields  play on the evolution of defect structures and grain boundaries, on
diffusive time scales. Examples for planar and circular grain boundaries
explain the essential atomistic processes and large scale simulations in 2D
are used to obtain statistical data on grain growth under the influence of
external fields.  \end{abstract}

\pacs{?}

\maketitle



\begin{figure}
\raisebox{-.5\height}{\includegraphics[width=0.35\textwidth]{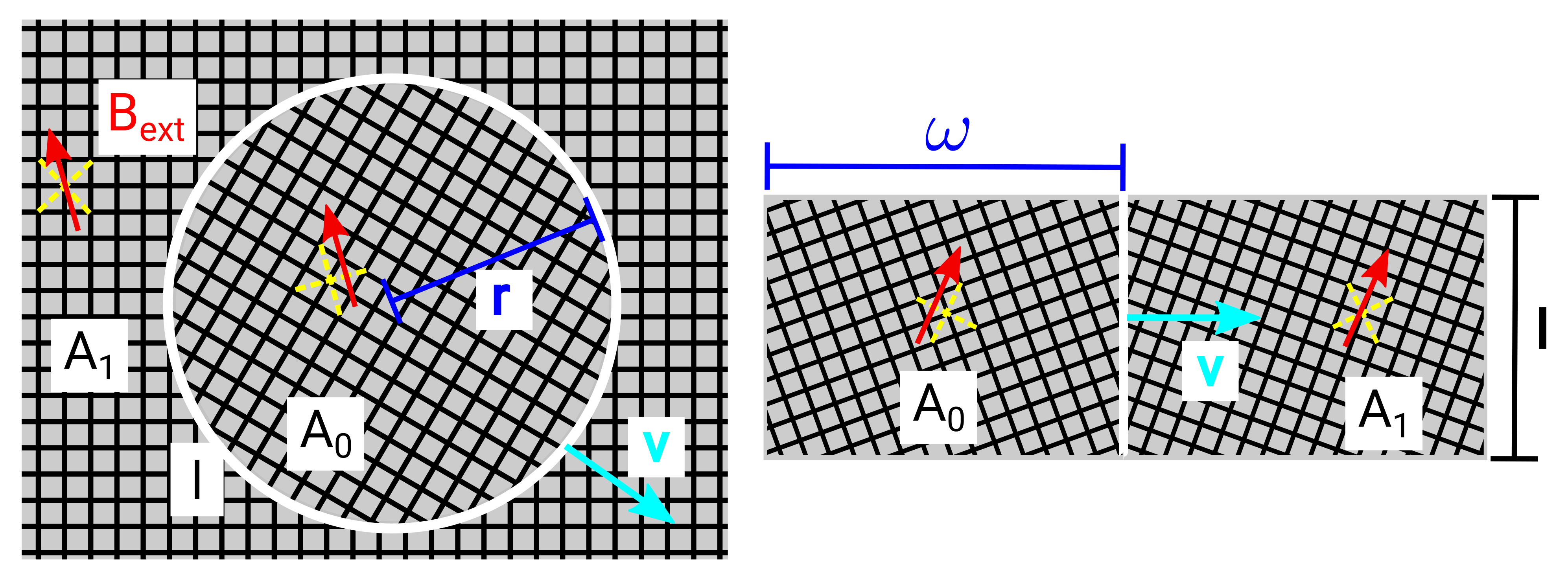}}
\caption{\label{fig:contMod}
Two grains with size A$_0$ and A$_1$ separated by a GB with size
$l$, which moves locally with the velocity $v$. The free energy
density in the bulk depends on the alignment of \BBext with the easy direction
of the crystal structure.} 
\end{figure}

It is well known that material properties of polycrystalline materials are 
strongly influenced by the average grain size.  For example, in some compounds 
the magnetic coercivity can increase by orders of magnitude as the grain size changes 
from nano to micron scales \cite{Herzer13,Chen03,Roy04,Xue08}. 
In metals the yield strength can not only change dramatically 
with grain size (the so-called Hall-Petch effect \cite{yip98,hall51,petch53,crack55,Lu90,chok89}) 
but it is also influenced by details of the grain size distribution \cite{Dahlberg14}.  
Each of the cases highlights the importance of the grain structure and the
technological need to understand and control its formation.  
The use of external magnetic fields offers additional degrees of freedom to
synthesize materials and to tailor the grain structure and thus material properties.
Although evidence for the interactions between external magnetic fields,
diffusion and irreversible deformation mechanisms have been gathered over the
years, see the review \cite{Guillonetal_MT_2018}, a global yet detailed
understanding of the interactions between magnetic fields and solid-state
matter transport is far from being reached. In this Letter we analyze the
properties of a theoretical model, which allows the description of the basic physics
of magnetocrystalline interactions in a multiscale approach, combining the
dynamics of defects, dislocation networks and grain boundaries with
experimentally accessible microstructure evolution on diffusive time scales.
The basic mechanisms of this interaction can be understood on thermodynamic 
arguments.  In magnetic materials the magnetic moments are aligned with a 
sufficiently strong external magnetic field. If the magnetic properties of the material are
anisotropic, the bulk free energy differs for differently oriented grains and
the energy difference can influence grain boundary (GB) movement. Assuming
setups of two differently oriented grains in a strong magnetic field, see
Fig~\ref{fig:contMod}, the total energy of the system reads $E=\gamma l +
\Delta f A_0 +f_1 (A_0+A_1)$, where $l$ is the length of GB and $A_i, f_i$ the
size and the energy density of the i-th grain, $\Delta f=f_0-f_1$ and
$\gamma$ the energy of the GB. The dynamics of the GB can be described by Mullins-type models
\cite{Mullins_JAP_1956}  
\begin{align}
\label{eq:vel}
v = -M \left( \gamma \kappa -\Delta f \right)
\end{align}
extended by the bulk energy difference \cite{AG89,TC94}, where $v$ is the 
normal velocity of the GB, $M$ a mobility function and
$\kappa$ the mean curvature. For a planar GB, $A_0 = w l$ and $\dot{A_0} =
\dot{w} l = v l =  M \Delta f l$. With $\dot{E}=\gamma  \dot{l} + \Delta
f \dot{A_0} $ we obtain  $\dot{E} =  M (\Delta f)^2 l$, a
constant normal velocity proportional to $\Delta f$ and a linearly decreasing
energy which scales with $(\Delta f)^2$. For a circular GB, $A_0 = \pi r^2$, we
obtain $\dot{E} = 2 \pi (\gamma + \Delta f r) \dot{r}$ and thus, equivalently
to classical nucleation theory, a critical grain size $r_c = -\gamma/\Delta f$, 
which leads to growth for a specific driving force in order
to decrease the energy. Both cases demonstrate the possibility to influence GB
movement by external magnetic fields.  However, this description ignores the 
underlying crystalline lattice which can influence the process.


It has been shown that
the complex dislocation structure along curved GB gives rise to a
misorientation-dependent mobility \cite{Winningetal_AM_2001}. Further studies
indicate that grain boundaries undergo thermal roughening associated with an
abrupt mobility change, leading to smooth (fast) and rough (slow) boundaries
\cite{Holmetal_Science_2010}, which can eventually lead to stagnation of the
growth process. The defect structure at triple junctions can lead to a
sufficiently small mobility limiting the rate of GB migration
\cite{Srinivasanetal_AM_1999,Upmanyuetal_AM_2002}. Also, tangential motion of
the lattices is possible. For low-angle GB, normal and tangential
motion are strongly coupled as a result of the geometric constraint that the
lattices of two crystals change continuously across the interface while the
GB moves \cite{Cahnetal_AM_2004}. As a consequence of this
coupling, grains rotate as they shrink, which leads to an increase in the GB 
energy per unit length, although the overall energy decreases since
the size of the boundary decreases \cite{Shanetal_Science_2004,Upmanyuetal_AM_2006,Trauttetal_AM_2012,Wuetal_AM_2012,hein14}. 
The phase field crystal (PFC) model \cite{Elderetal_PRL_2002,Elderetal_PRE_2004,Elderetal_PRE_2007,Teeffelenetal_PRE_2009},
captures all these complex features and numerical simulations of the model 
have been shown to recover the characteristic grain size distribution in 
agreement with detailed experimental results \cite{BBV14}.
Numerous publications have shown the model to capture the essential physics
of atomic-scale elastic and plastic effects that accompany diffusive phase
transformations, such as solidification, dislocation kinetics and solid-state
precipitation, see \cite{Emmerichetal_AP_2012} for a review. 
In \cite{FPK13} the model is coupled with magnetization to generate 
a ferromagnetic solid below a density-dependent Curie temperature. In \cite{SSP15} this model 
is extended and used to demonstrate the influence of magnetic fields on the growth of
crystal grains. These results indicate that a greater portion of grains
evolve to become aligned along the easy direction of the crystal structure with
respect to the orientation of the external magnetic field.  
We here use it to predict the influence of the magnetic field on grain 
coarsening in polycrystals.  Consistent with the thermodynamic arguments 
we find that when the magnetic field is applied, the average grain size increases and 
the number of grain along the easy direction with respect to the field increases. 
However, it is also found that the grains become elongated when the field is 
applied. The elongation occurs due to an anisotropic GB mobility in the 
presence of an applied field. Details of the study are presented below.

 
The model \cite{FPK13,SSP15} combines a PFC model for crystalline ordering in
terms of the rescaled number density \phii with a mean field approximation for
the averaged magnetization \mm. The energy consists of three contributions,
\ffpfc $\,$  related to the local ordering of the crystal, \ffm $\,$ related to
the local orientation of the magnetic moment and \ffcoup $\,$ related to the coupling 
between crystal structure and magnetization and reads:
$\FF[\phii,\mm] = \int \ffpfc[\phii] + \omega_B \ffm[\mm] + \omega_B \ffcoup[\phii,\mm] \ddr $  
with
\begin{eqnarray*}
\ffpfc[\phii]  &=&  \frac{1}{2} \! \phii[\rr]^2 \!-\! \frac{t}{6} \! \phii[\rr]^3 \!+\! \frac{v}{12} \! \phii[\rr]^4 \\
&&\quad -\frac{1}{2}  \phii[\rr] \int C_2(\rr-\rr') \phii[\rr'] \ddr'  \\
\ffm[\mm]  &=&  \frac{W_0^2}{2} \! \left( \nabla \! \cdot \! \mm\right)^2 
 \! + \! r_m  \frac{\mm^2}{2} \! + \! \gamma_m \frac{\mm^4}{4} \! - \! \mm \cdot \BB \! + \! \frac{\BB^2}{2} \\
 \ffcoup[\phii,\mm]   &=&   - \omega_m \phii^2  \frac{\mm^2}{2} \! - \!  \sum_{j=1}^2 
\frac{\alpha_{2j}}{2 j} \left( \mm \cdot \nabla \phii \right) ^{2j}, 
\end{eqnarray*}
where $\omega_B$ is a parameter to control the influence of the magnetic energy. 
In order to maximize
the anisotropy in the 2D setting, a square ordering of the crystal is
preferred, which is realized within the XPFC formulation for $\ffpfc[\phii]$, 
see \cite{GPR10,OSP13} and SI.

Magnetization in an isotropic and homogenous material is modeled by
$\ffm[\mm]$. The first three terms define a mean field theory of a vector field
which is minimized by $\mm=0$ for $r_m > 0$ and
$\mm=-r_m/\gamma_m$ for $r_m<0$. Thus, a negative $r_m$ leads to
ferromagnetic properties. The last two terms describe the interaction of the
magnetization with an external and a self-induced magnetic field, $\BBext$ and
$\BBind$, respectively.  The magnetic field is defined as $\BB=\BBext+\BBind$,
where \BBind is defined with help of the vector potential: $\BBind=\nabla
\times \VA $ and $ \nabla^2 \VA = - \nabla \times \mm$. The anisotropy of the
material is due to the crystalline structure of the material. Thus, the
magnetization has to depend on the local structure represented by \phii and
vice versa. The first term in $\ffcoup[\phii,\mm]$, changes the ferromagnetic
transition in the magnetic free energy. On average $\phii^2$ is larger in the
crystal than in the homogeneous phase. Thus, $\omega_m$ and $r_m$ can be chosen
to realize a paramagnetic homogeneous phase and a ferromagnetic crystal. The
second term depends on average on the relative orientation of the crystalline
structure with respect to the magnetization. In our case, it lead to an energetic
minimum if the magnetization is aligned with the diagonal of the square
crystal. Thus, the easy directions of magnetization are along the
\hkl<1 1>-directions. The number density \phii evolve according to conserved
dynamics and magnetization according to non-conserved dynamics, 
\begin{align}
\!\!\frac{\partial \phii}{\partial t} = M_n \nabla^2 \frac{\delta \FF[\phii,\mm]}{\delta \phii}, \quad
\frac{\partial \mmi}{\partial t} = - M_m \frac{\delta
\FF[\phii,\mm]}{\delta \mmi}
\end{align}
$i = 1,2$, respectively. See SI for details.

To measure the magnetic anisotropy we consider a single crystal and vary
\BBext. The simulation domain perfectly fits the equilibrium crystal for
\BBext=0 and is small enough to prevent the appearance of magnetic domains. The
parameters are chosen for a ferromagnetic material, see SI for details.
\begin{figure}[htb]
\includegraphics[width=0.36\textwidth]{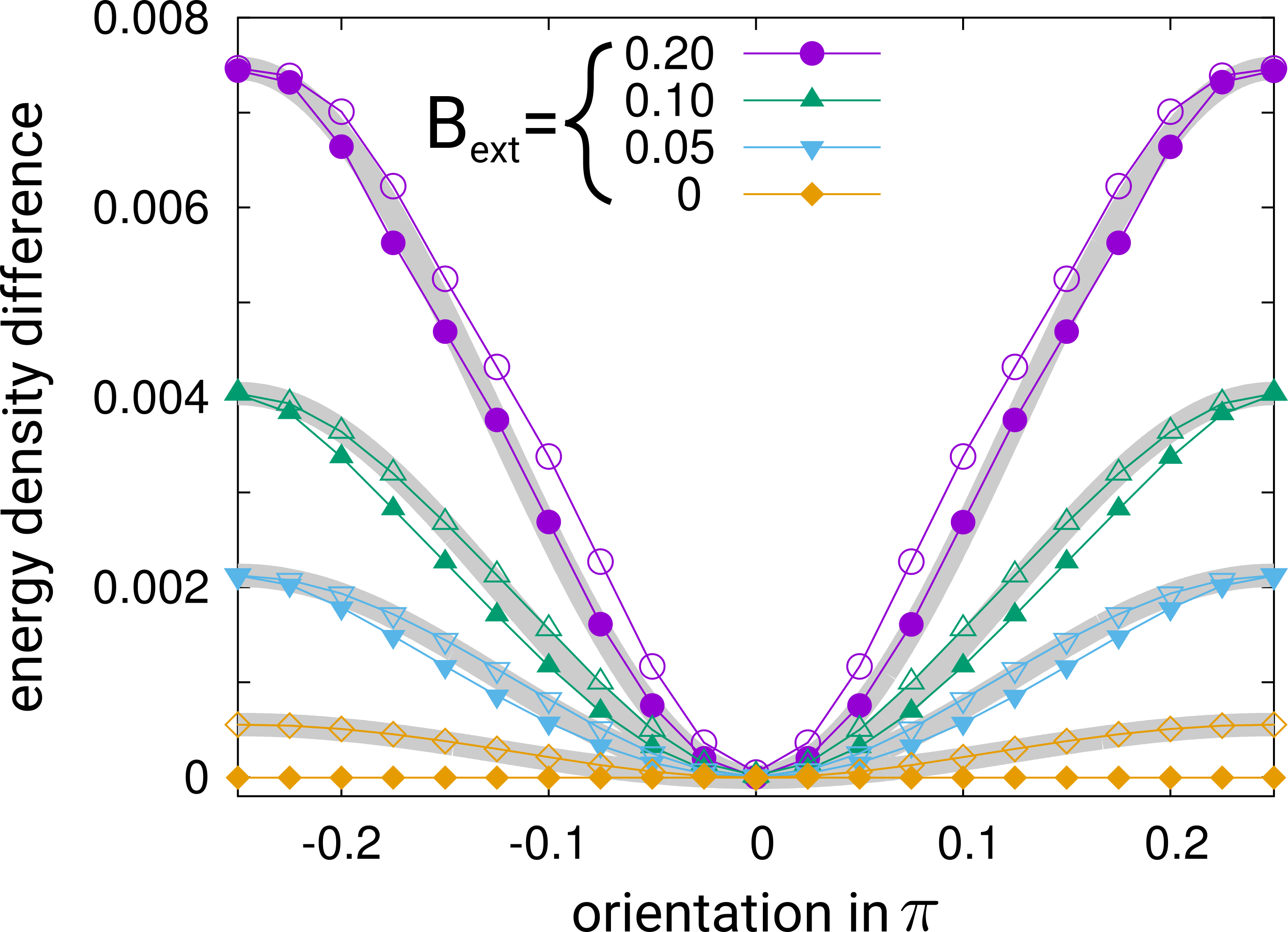}
\caption{\label{fig:aniso} Energy density deviation in a single crystal induced
by \BBext and measured relative to a crystal preferably aligned with \BBext. 
The orientation with respect to the crystal structure and strength of \BBext
is varied. Open symbols correspond to forced alignment of
magnetic moments with \BBext, closed symbols show computed magnetic moments,
gray curves show fits by cosine-functions.} 
\end{figure}
\begin{figure*}[hbt]
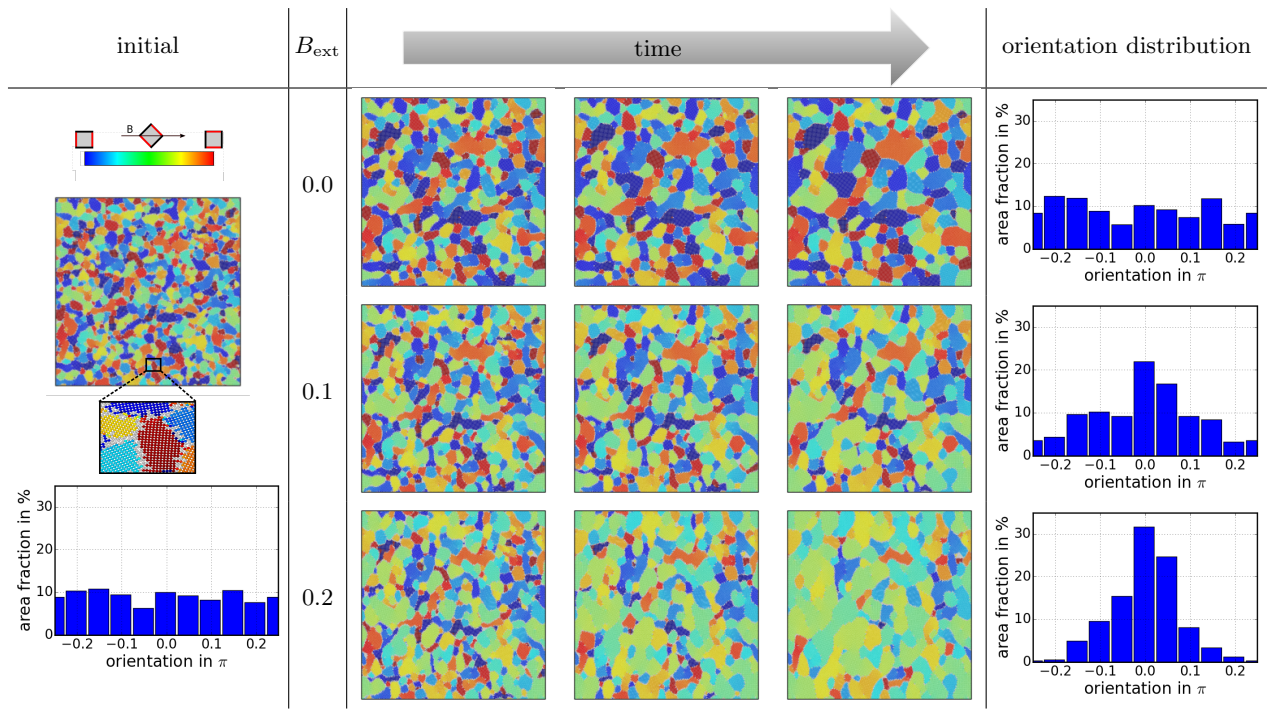

\begin{tabular}{c|c|ccc|c}
initial & \NBext  & \multicolumn{3}{c|}{\tikzfancyarrow[7cm]{time} }& orientation distribution  \\ \hline
\multirow{3}{*}[2em]{\shortstack{\includegraphics*[width =2cm] {{{./figures/colorMapOrientNice}}}
\\ \includegraphics*[width =0.15\textwidth] {{{./figures/coarsInitAndZoom}}}
\\ \includegraphics*[width =0.2\textwidth] {{{./figures/initOrient}}} 
}}
 & 0.0 & 
\raisebox{-.5\height}{\includegraphics*[width =0.15\textwidth] {{{./figures/coarsB0_25}}}}& 
\raisebox{-.5\height}{\includegraphics*[width =0.15\textwidth] {{{./figures/coarsB0_50}}}} & 
\raisebox{-.5\height}{\includegraphics*[width =0.15\textwidth] {{{./figures/coarsB0_100}}}} & 
\raisebox{-.5\height}{\includegraphics*[width =0.2\textwidth] {{{./figures/B00Orient}}}} \\
 & 0.1 & 
\raisebox{-.5\height}{\includegraphics*[width =0.15\textwidth] {{{./figures/coarsB01_25}}}}& 
\raisebox{-.5\height}{\includegraphics*[width =0.15\textwidth] {{{./figures/coarsB01_50}}}} & 
\raisebox{-.5\height}{\includegraphics*[width =0.15\textwidth] {{{./figures/coarsB01_96}}}} & 
\raisebox{-.5\height}{\includegraphics*[width =0.2\textwidth] {{{./figures/B01Orient}}}} \\
 & 0.2 & 
\raisebox{-.5\height}{\includegraphics*[width =0.15\textwidth] {{{./figures/coarsB02_25}}}}& 
\raisebox{-.5\height}{\includegraphics*[width =0.15\textwidth] {{{./figures/coarsB02_50}}}} & 
\raisebox{-.5\height}{\includegraphics*[width =0.15\textwidth] {{{./figures/coarsB02_100}}}} & 
\raisebox{-.5\height}{\includegraphics*[width =0.2\textwidth] {{{./figures/B02Orient}}}} \\
\end{tabular}
\caption{\label{fig:coars} (left) Initial configuration for coarsening
simulation. The color shows the local orientation of the crystal with respect to the
external magnetic field. The direction of the external magnetic field is in
x-direction and corresponds to grains oriented in the easy direction (green).
For the inlet the maxima of $\phii$ are visualized as atoms. The orientation
distribution is isotropic. (middle) Coarsening simulation for different \BBext
(up-down) with snapshots in time (left-right). (right) Orientation distribution
at final time of coarsening process. For the used parameters see Si. The
computational domain is $409.6 \times 409.6$.} 
\end{figure*}
Fig.~\ref{fig:aniso} shows the anisotropy of the bulk free energy with respect to the
orientation of the magnetic moments with and without an external magnetic
field.  Restricting the magnetic moments to the direction of the external
magnetic field, leads to slightly larger bulk energies for orientations not
along hard and easy direction. This is due to the reduced degrees of freedom
for energy minimization and shows that in the full model in these cases the
magnetic moments are not perfectly aligned with \BBext. However, the
differences are small. The magnetic anisotropy for both cases follows the
4-fold symmetry of the crystal and the easy directions are along the
\hkl<1 1>-direction. It can be approximated by a cosine (shaded line). Increasing
\BBext increases the anisotropy as well as the mean magnetization. The model
also includes magnetostriction effects \cite{FPK13}. The crystal slightly tends
to elongate along the easy direction aligned with \BBext, see SI for details.

To show the impact of external magnetic fields on the texture evolution during
coarsening we prepared a polycrystalline sample, see Fig.~\ref{fig:coars}. 
An initially randomly perturbated density field is evolved without magnetic interaction until the fine polycrystalline structure appears. 
Any particle with four neighbors is identified
as a particle in a crystalline structure and the local orientation of the
crystal with respect to the external magnetic field is calculated and visualized. 
Starting from this initial condition the evolution equations are solved with 
small random magnetization for different external magnetic fields, applied in x-direction. 
For $\NBext=0$
there is no energetically preferred orientation and coarsening is only due
to minimization of GB energy. Small grains vanish and larger grains grow. The
average grain size increases and the orientation distribution stays isotropic.
Applying an external field leads to a preferred growth of grains which are
aligned preferably with respect to the external magnetic field, the easy direction
(green). Thus, the not aligned grains (blue and red) vanish and the orientation
distribution peaks near the aligned grain orientation.  This is in qualitative
agreement with experiments, e.g. on Zn and Ti sheets \cite{Molodovetal_SM_2006}, 
and classical grain growth simulations of Mullins type with an analytical
magnetic driving force \cite{Barrales-Moraetal_CMS_2007}. The additional
driving force, due to the external magnetic field, also enhances the coarsening
process, which can already be seen by comparing the final textures in
Fig.~\ref{fig:coars} and which has also been observed experimentally, e.g.
during annealing of FeCo under high steady magnetic fields
\cite{Rivoirard_JOM_2013}. Increasing $\BBext$ leads to more pronounced grain
orientation selection. For further quantification of these effects, see SI. 

\begin{figure}
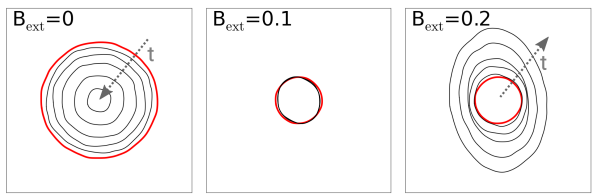

\begin{tabular}{ccc}
\raisebox{-.5\height}{\includegraphics*[width =0.14\textwidth]{{{figures/B00_isoN}}}} &
\raisebox{-.5\height}{\includegraphics*[width =0.14\textwidth]{{{figures/B01_iso}}}} &
\raisebox{-.5\height}{\includegraphics*[width =0.14\textwidth]{{{figures/B02_isoN}}}} \\
\end{tabular}
\caption{\label{fig:sgGrain} A circular grain embedded in a matrix (red
isoline). The external magnetic filed is aligned with the easy direction of the
circular grain. Dependent on strength of \BBext the grain shrinks, stagnates or
grows, see SI for details. } \end{figure}

In order to analyze these results in more detail we consider the two simple
examples illustrated in Fig.~\ref{fig:contMod}. We start with a rotated crystal
embedded in a matrix, see Fig.~\ref{fig:sgGrain}. For $\NBext=0$ the grain
shrinks and vanishes in order to minimize GB energy. \BBext aligned with the
easy direction of the rotated grain induces an opposite driving force, which
for $\NBext=0.1$ balances the GB energy, while increasing $\BBext$ above this
threshold leads to growth of the grain. This is in accordance with the
continuous description. However, for $\NBext=0.2$ the evolution is anisotropic,
first a square like shape is reached, resampling the 4-fold crystalline
symmetry, while further growth breaks this symmetry, the grain becomes
elongated perpendicular to $\BBext$. This may be explained by thermodynamic or
kinetic reasons \cite{Sek05,Hanetal_PMS_2018}. 

\begin{figure}
\includegraphics[width=0.4\textwidth]{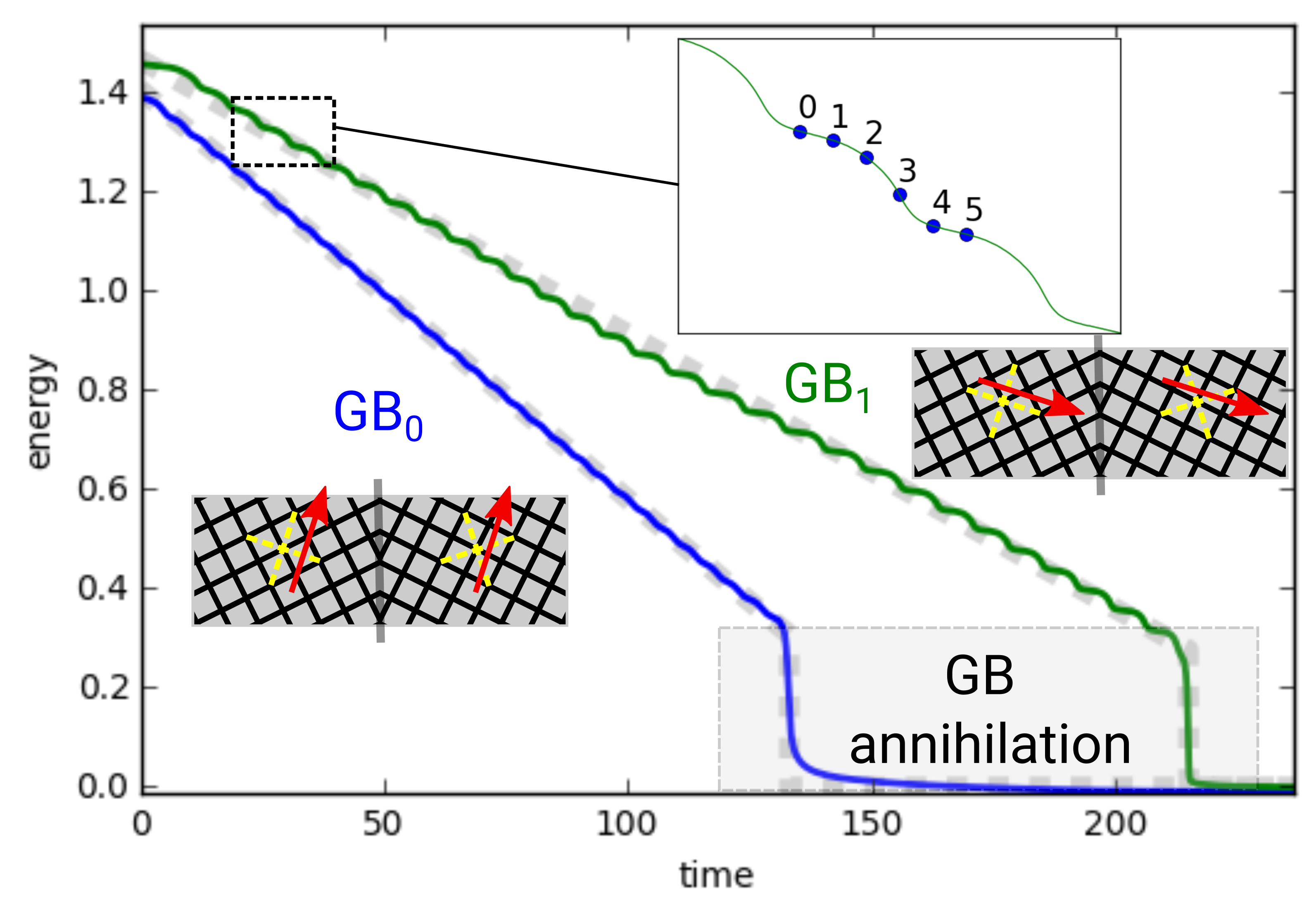}
\caption{\label{fig:planarGB} Two setups of a symmetric tilt GB in
a periodic domain, $\NBext = 0.1$ is alined with the easy directions of the
left grain. Both setups lead to the same driving force, but the energy decay
differs. } 
\end{figure}

\begin{figure}[htb]
\begin{tabular}{ccc}
0 & 1 & 2\\
\raisebox{-.5\height}{\includegraphics*[width=0.15\textwidth]{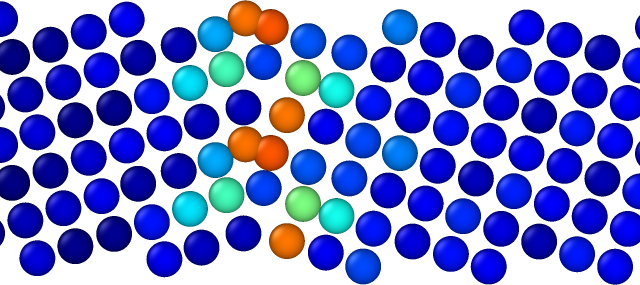}} & 
\raisebox{-.5\height}{\includegraphics*[width=0.15\textwidth]{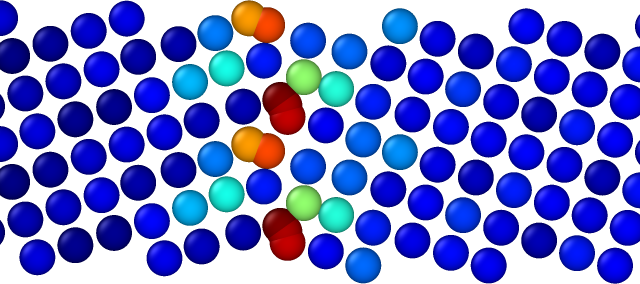}} &
\raisebox{-.5\height}{\includegraphics*[width=0.15\textwidth]{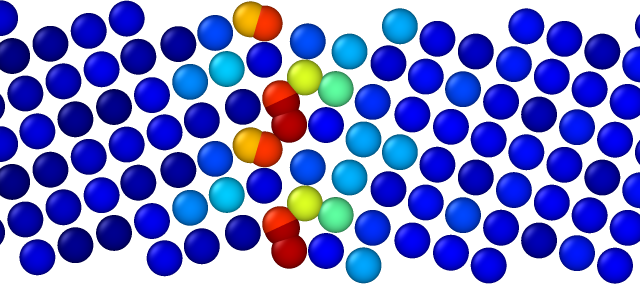}} \\
3 & 4 & 5 \\
\raisebox{-.5\height}{\includegraphics*[width=0.15\textwidth]{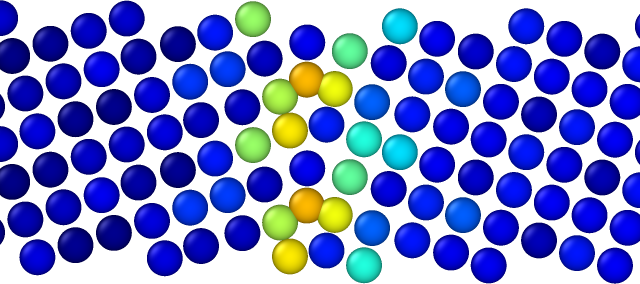}} &
\raisebox{-.5\height}{\includegraphics*[width=0.15\textwidth]{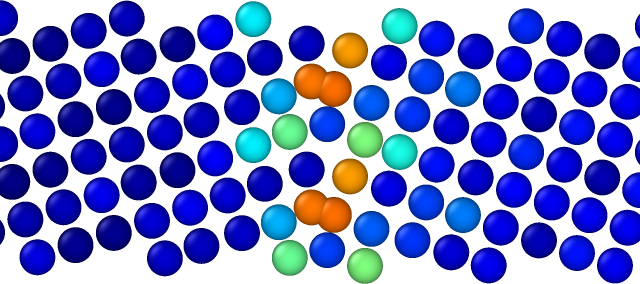}} &
\raisebox{-.5\height}{\includegraphics*[width=0.15\textwidth]{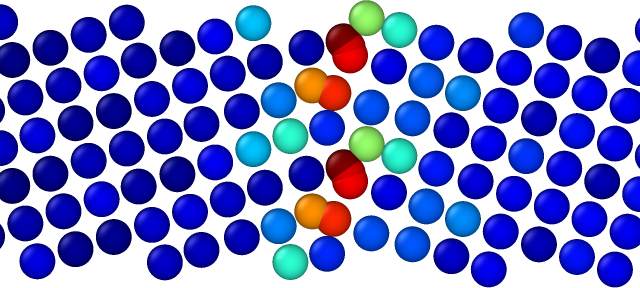}} 
\end{tabular}
\begin{center}
\caption{\label{fig:step} Particle picture of the GB during evolution over one
unit length. The particles are located according to maxima in the density field
$\phii$. The color is the energy density at the position of the particle and
serves as a measure of the local energy, see \cite{Koehleretal_PRL_2017}.
During the slow evolution (0-2) the energy of the particles at the GB increases
until the energy barrier is overcome by the magnetic driving force leading to a
speed up of the GB and a decrease of the energy at the GB (2-3), before the
next barrier is reached (3-4) and the energy at the GB increases again (4-5).
}
\end{center}
\end{figure}

Within the continuous description of eq. \eqref{eq:vel} the shape reached 
for $\NBext=0.2$ requires either the GB energy $\gamma$ parallel to \BBext to 
be roughly twice the energy perpendicular to $\BBext$ or the mobility $M$ of
parallel and perpendicular GB has to vary by a factor of two or some combination 
of both.  To separate thermodynamic ($\gamma$) and kinetic effects ($M$) of GB movement, 
we consider a planar GB. According to the continuum description the velocity of
the planar GB is proportional to the driving force $\Delta f$. Thus, the decay
of total energy is linear and the mobility can be extracted, $M= -v/\Delta f$.
To maximize the influence of \BBext two symmetric high
angle GB are placed in an elongated periodic domain. 
\BBext is aligned with the easy direction of the left grain. Due to symmetry
the magnetic field can be rotated  by $\pi/2$. In one situation the
magnetization is more aligned and in the other more perpendicular to the GB, see
Fig. \ref{fig:planarGB}, which shows the setup and the energy decay for both
situations. 
 The initial condition is achieved by a purely structural relaxation
with $\omega_B=0$. Then the coupling with \BBext is switched on. After some
initial reconfiguration, which adjusts the density field $\phii$, the energy
decays on average linearly. The GBs move with constant speed reducing the size
of the grain not aligned with $\BBext$ until they vanish. The final
annihilation of the GB leads to a sudden drop in energy, which is proportional
to $\gamma$ and equal in both cases. However, the energy decays faster in the
case of a more aligned \BBext with the GB, implying faster GB velocity and 
in turn a larger GB mobility. 
A closer look at the energy decay shows a step like function. This
reflects the crystalline structure of the GB. In order to move the
GB by a unit length it has to pass some energetically unfavorable positions, see
Fig.~\ref{fig:step} and SI for details.  
\begin{figure}
\includegraphics[width=0.36\textwidth]{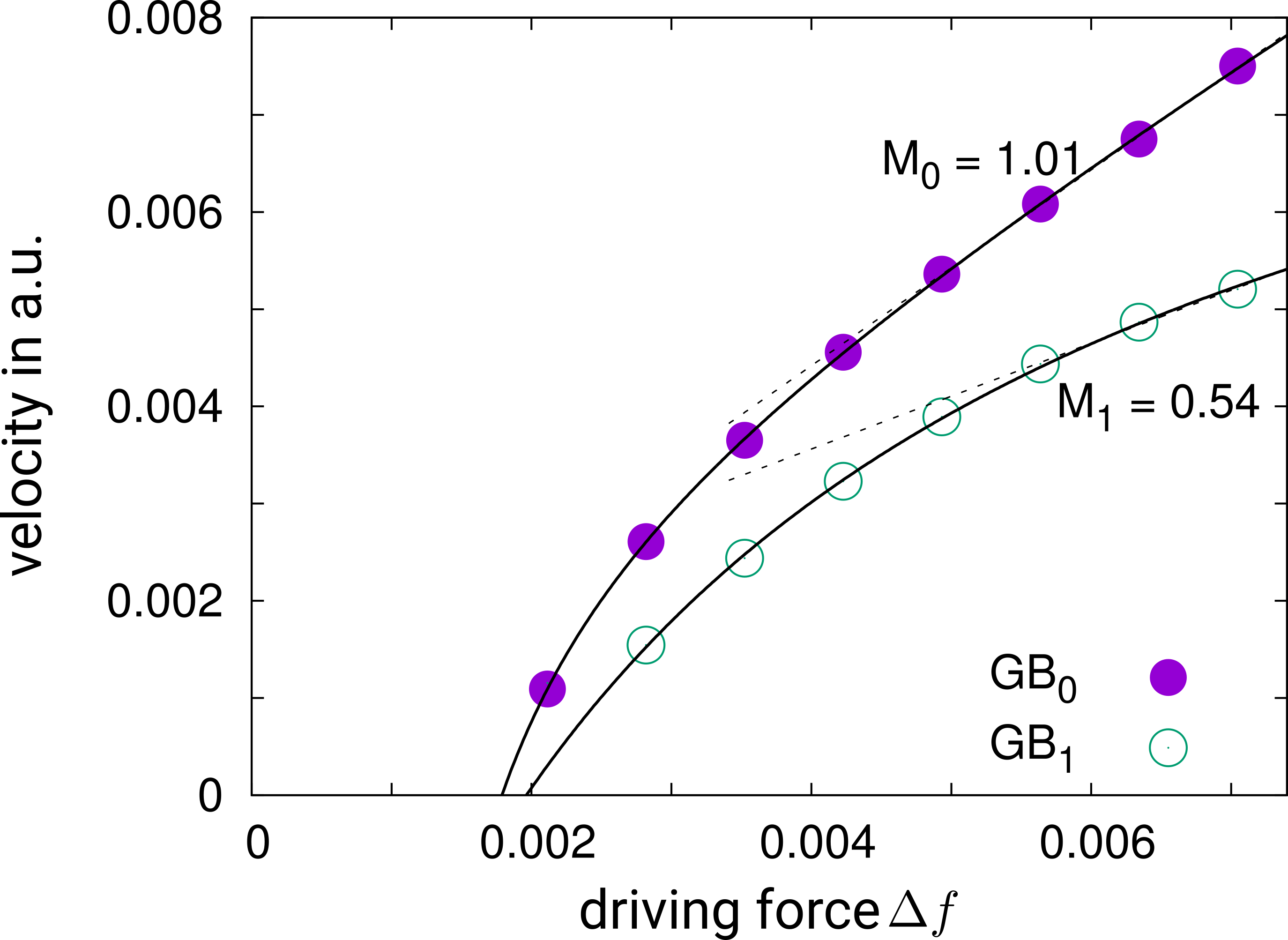}
\caption{\label{fig:velocity} Velocity extracted for the setups defined in
Fig.~\ref{fig:planarGB}. For small external magnetic field the GB is pinned and
does not move at all. High driving forces lead to a linear increase of velocity
with $\Delta f$ and an assumed mobility becomes constant. The mobility differs 
by a factor of two.}
\end{figure}
Varying the magnitude of $\BBext$ changes the driving force and the
velocity of the GB, see Fig.~\ref{fig:velocity}. For large driving forces the
dependency of the velocity is linear for both cases but by a factor two smaller
for the case of $\BBext$ more perpendicular to the GB. For a driving forces
below a threshold the GB does not move, indicating the presence of an
activation barrier, which has also been measured experimentally for planar GB
in Zn bicrystals \cite{Guensteretal_SM_2010}. For intermediate regimes the
mobility increases.  As a consequence, the anisotropy seen in
Fig.~\ref{fig:sgGrain} can be attributed to kinetics and not thermodynamic
effects, which was also claimed in \cite{Molodovetal_SM_2006} by interpreting
the experiments on Zn and Ti. 

	In summary we have shown that an applied magnetic field can 
increase the coarsening rate in grain growth processes, due to the 
lower energy of grains with their easy axis in line with the 
applied field.  We have also shown that the mobility of GB is 
anisotropic with respect to the applied magnetic 
field. This kinetic effect leads to elongated grains.  
Both of these influences are intimately related to the 
magnetically anisotropic nature of the model studied.  
That is, the crystal reacts elastically on applied magnetic fields 
(magnetorestriction) and additionally changes in the density field 
reflecting the two fold symmetry of \BBext may lead to preferred 
diffusion path and, thus, influence the mobility.   
It should be noted that the study examined the influence of 
an applied field on a ferromagnetic nano-crystalline system
and did not examine the influence of magnetic field on the 
initial nucleation stage. This is left for future study.

\begin{acknowledgments}
AV and RB acknowledge support by the German Research Foundation (DFG) under grant SPP1959. We further acknowledge computing resources provided at J\"ulich Supercomputing Center under grant HDR06.
\end{acknowledgments}

\bibliography{MagneticGrainBoundary}

\begin{thebibliography}{42}%
\makeatletter
\providecommand \@ifxundefined [1]{%
 \@ifx{#1\undefined}
}%
\providecommand \@ifnum [1]{%
 \ifnum #1\expandafter \@firstoftwo
 \else \expandafter \@secondoftwo
 \fi
}%
\providecommand \@ifx [1]{%
 \ifx #1\expandafter \@firstoftwo
 \else \expandafter \@secondoftwo
 \fi
}%
\providecommand \natexlab [1]{#1}%
\providecommand \enquote  [1]{``#1''}%
\providecommand \bibnamefont  [1]{#1}%
\providecommand \bibfnamefont [1]{#1}%
\providecommand \citenamefont [1]{#1}%
\providecommand \href@noop [0]{\@secondoftwo}%
\providecommand \href [0]{\begingroup \@sanitize@url \@href}%
\providecommand \@href[1]{\@@startlink{#1}\@@href}%
\providecommand \@@href[1]{\endgroup#1\@@endlink}%
\providecommand \@sanitize@url [0]{\catcode `\\12\catcode `\$12\catcode
  `\&12\catcode `\#12\catcode `\^12\catcode `\_12\catcode `\%12\relax}%
\providecommand \@@startlink[1]{}%
\providecommand \@@endlink[0]{}%
\providecommand \url  [0]{\begingroup\@sanitize@url \@url }%
\providecommand \@url [1]{\endgroup\@href {#1}{\urlprefix }}%
\providecommand \urlprefix  [0]{URL }%
\providecommand \Eprint [0]{\href }%
\providecommand \doibase [0]{http://dx.doi.org/}%
\providecommand \selectlanguage [0]{\@gobble}%
\providecommand \bibinfo  [0]{\@secondoftwo}%
\providecommand \bibfield  [0]{\@secondoftwo}%
\providecommand \translation [1]{[#1]}%
\providecommand \BibitemOpen [0]{}%
\providecommand \bibitemStop [0]{}%
\providecommand \bibitemNoStop [0]{.\EOS\space}%
\providecommand \EOS [0]{\spacefactor3000\relax}%
\providecommand \BibitemShut  [1]{\csname bibitem#1\endcsname}%
\let\auto@bib@innerbib\@empty
\bibitem [{\citenamefont {Herzer}(2013)}]{Herzer13}%
  \BibitemOpen
  \bibfield  {author} {\bibinfo {author} {\bibfnamefont {G.}~\bibnamefont
  {Herzer}},\ }\href@noop {} {\bibfield  {journal} {\bibinfo  {journal} {Acta
  Mater.}\ }\textbf {\bibinfo {volume} {61}},\ \bibinfo {pages} {718} (\bibinfo
  {year} {2013})}\BibitemShut {NoStop}%
\bibitem [{\citenamefont {Chen}\ \emph {et~al.}(2003)\citenamefont {Chen},
  \citenamefont {Kodat}, \citenamefont {Walmer}, , \citenamefont {Chen},
  \citenamefont {Willard},\ and\ \citenamefont {Harris}}]{Chen03}%
  \BibitemOpen
  \bibfield  {author} {\bibinfo {author} {\bibfnamefont {C.~H.}\ \bibnamefont
  {Chen}}, \bibinfo {author} {\bibfnamefont {S.}~\bibnamefont {Kodat}},
  \bibinfo {author} {\bibfnamefont {M.~H.}\ \bibnamefont {Walmer}}, , \bibinfo
  {author} {\bibfnamefont {S.-F.}\ \bibnamefont {Chen}}, \bibinfo {author}
  {\bibfnamefont {M.~A.}\ \bibnamefont {Willard}}, \ and\ \bibinfo {author}
  {\bibfnamefont {V.~G.}\ \bibnamefont {Harris}},\ }\href@noop {} {\bibfield
  {journal} {\bibinfo  {journal} {J. App. Phys.}\ }\textbf {\bibinfo {volume}
  {93}},\ \bibinfo {pages} {7966} (\bibinfo {year} {2003})}\BibitemShut
  {NoStop}%
\bibitem [{\citenamefont {Roy}\ \emph {et~al.}(2004)\citenamefont {Roy},
  \citenamefont {Dubenko}, \citenamefont {Edorh},\ and\ \citenamefont
  {Ali}}]{Roy04}%
  \BibitemOpen
  \bibfield  {author} {\bibinfo {author} {\bibfnamefont {S.}~\bibnamefont
  {Roy}}, \bibinfo {author} {\bibfnamefont {I.}~\bibnamefont {Dubenko}},
  \bibinfo {author} {\bibfnamefont {D.~D.}\ \bibnamefont {Edorh}}, \ and\
  \bibinfo {author} {\bibfnamefont {N.}~\bibnamefont {Ali}},\ }\href@noop {}
  {\bibfield  {journal} {\bibinfo  {journal} {J. App. Phys.}\ }\textbf
  {\bibinfo {volume} {96}},\ \bibinfo {pages} {1202} (\bibinfo {year}
  {2004})}\BibitemShut {NoStop}%
\bibitem [{\citenamefont {Xue}\ \emph {et~al.}(2008)\citenamefont {Xue},
  \citenamefont {Chai}, \citenamefont {Li},\ and\ \citenamefont {Fan}}]{Xue08}%
  \BibitemOpen
  \bibfield  {author} {\bibinfo {author} {\bibfnamefont {D.}~\bibnamefont
  {Xue}}, \bibinfo {author} {\bibfnamefont {G.}~\bibnamefont {Chai}}, \bibinfo
  {author} {\bibfnamefont {X.}~\bibnamefont {Li}}, \ and\ \bibinfo {author}
  {\bibfnamefont {X.}~\bibnamefont {Fan}},\ }\href@noop {} {\bibfield
  {journal} {\bibinfo  {journal} {J. Mag. Mag. Mat.}\ }\textbf {\bibinfo
  {volume} {320}},\ \bibinfo {pages} {1541} (\bibinfo {year}
  {2008})}\BibitemShut {NoStop}%
\bibitem [{\citenamefont {Yip}(1998)}]{yip98}%
  \BibitemOpen
  \bibfield  {author} {\bibinfo {author} {\bibfnamefont {S.}~\bibnamefont
  {Yip}},\ }\href@noop {} {\bibfield  {journal} {\bibinfo  {journal} {Nature}\
  }\textbf {\bibinfo {volume} {391}},\ \bibinfo {pages} {532} (\bibinfo {year}
  {1998})}\BibitemShut {NoStop}%
\bibitem [{\citenamefont {Hall}(1951)}]{hall51}%
  \BibitemOpen
  \bibfield  {author} {\bibinfo {author} {\bibfnamefont {E.~O.}\ \bibnamefont
  {Hall}},\ }\href@noop {} {\bibfield  {journal} {\bibinfo  {journal} {Proc.
  Phys. Soc. London, Sect. B}\ }\textbf {\bibinfo {volume} {64}},\ \bibinfo
  {pages} {747} (\bibinfo {year} {1951})}\BibitemShut {NoStop}%
\bibitem [{\citenamefont {Petch}(1953)}]{petch53}%
  \BibitemOpen
  \bibfield  {author} {\bibinfo {author} {\bibfnamefont {N.~J.}\ \bibnamefont
  {Petch}},\ }\href@noop {} {\bibfield  {journal} {\bibinfo  {journal} {J. Iron
  Steel Inst., London}\ }\textbf {\bibinfo {volume} {174}},\ \bibinfo {pages}
  {25} (\bibinfo {year} {1953})}\BibitemShut {NoStop}%
\bibitem [{\citenamefont {Cracknell}\ and\ \citenamefont
  {Petch}(1955)}]{crack55}%
  \BibitemOpen
  \bibfield  {author} {\bibinfo {author} {\bibfnamefont {A.}~\bibnamefont
  {Cracknell}}\ and\ \bibinfo {author} {\bibfnamefont {N.}~\bibnamefont
  {Petch}},\ }\href@noop {} {\bibfield  {journal} {\bibinfo  {journal} {Acta
  Metall.}\ }\textbf {\bibinfo {volume} {3}},\ \bibinfo {pages} {186} (\bibinfo
  {year} {1955})}\BibitemShut {NoStop}%
\bibitem [{\citenamefont {Lu}\ \emph {et~al.}(1990)\citenamefont {Lu},
  \citenamefont {Wei},\ and\ \citenamefont {Wang}}]{Lu90}%
  \BibitemOpen
  \bibfield  {author} {\bibinfo {author} {\bibfnamefont {K.}~\bibnamefont
  {Lu}}, \bibinfo {author} {\bibfnamefont {W.}~\bibnamefont {Wei}}, \ and\
  \bibinfo {author} {\bibfnamefont {J.}~\bibnamefont {Wang}},\ }\href@noop {}
  {\bibfield  {journal} {\bibinfo  {journal} {Scr. Metall. Mater.}\ }\textbf
  {\bibinfo {volume} {24}},\ \bibinfo {pages} {2319} (\bibinfo {year}
  {1990})}\BibitemShut {NoStop}%
\bibitem [{\citenamefont {Chokshi}\ \emph {et~al.}(1989)\citenamefont
  {Chokshi}, \citenamefont {Rosen}, \citenamefont {Karch},\ and\ \citenamefont
  {Gleiter}}]{chok89}%
  \BibitemOpen
  \bibfield  {author} {\bibinfo {author} {\bibfnamefont {A.}~\bibnamefont
  {Chokshi}}, \bibinfo {author} {\bibfnamefont {A.}~\bibnamefont {Rosen}},
  \bibinfo {author} {\bibfnamefont {J.}~\bibnamefont {Karch}}, \ and\ \bibinfo
  {author} {\bibfnamefont {H.}~\bibnamefont {Gleiter}},\ }\href@noop {}
  {\bibfield  {journal} {\bibinfo  {journal} {Scr. Metall.}\ }\textbf {\bibinfo
  {volume} {23}},\ \bibinfo {pages} {1679} (\bibinfo {year}
  {1989})}\BibitemShut {NoStop}%
\bibitem [{\citenamefont {Dahlberg}\ and\ \citenamefont
  {Faleskog}()}]{Dahlberg14}%
  \BibitemOpen
  \bibfield  {author} {\bibinfo {author} {\bibfnamefont {C.~F.}\ \bibnamefont
  {Dahlberg}}\ and\ \bibinfo {author} {\bibfnamefont {J.}~\bibnamefont
  {Faleskog}},\ }\href@noop {} {\ }\BibitemShut {NoStop}%
\bibitem [{\citenamefont {Guillon}\ \emph {et~al.}(2018)\citenamefont
  {Guillon}, \citenamefont {Els\"asser}, \citenamefont {Gutfleisch},
  \citenamefont {Janek}, \citenamefont {Korte-Kerzel}, \citenamefont {Raabe},\
  and\ \citenamefont {Volkert}}]{Guillonetal_MT_2018}%
  \BibitemOpen
  \bibfield  {author} {\bibinfo {author} {\bibfnamefont {O.}~\bibnamefont
  {Guillon}}, \bibinfo {author} {\bibfnamefont {C.}~\bibnamefont {Els\"asser}},
  \bibinfo {author} {\bibfnamefont {O.}~\bibnamefont {Gutfleisch}}, \bibinfo
  {author} {\bibfnamefont {J.}~\bibnamefont {Janek}}, \bibinfo {author}
  {\bibfnamefont {S.}~\bibnamefont {Korte-Kerzel}}, \bibinfo {author}
  {\bibfnamefont {D.}~\bibnamefont {Raabe}}, \ and\ \bibinfo {author}
  {\bibfnamefont {C.}~\bibnamefont {Volkert}},\ }\href@noop {} {\bibfield
  {journal} {\bibinfo  {journal} {Mater. Today}\ }\textbf {\bibinfo {volume}
  {21}},\ \bibinfo {pages} {527} (\bibinfo {year} {2018})}\BibitemShut
  {NoStop}%
\bibitem [{\citenamefont {Mullins}(1956)}]{Mullins_JAP_1956}%
  \BibitemOpen
  \bibfield  {author} {\bibinfo {author} {\bibfnamefont {W.}~\bibnamefont
  {Mullins}},\ }\href@noop {} {\bibfield  {journal} {\bibinfo  {journal} {J.
  Appl. Phys.}\ }\textbf {\bibinfo {volume} {27}},\ \bibinfo {pages} {900}
  (\bibinfo {year} {1956})}\BibitemShut {NoStop}%
\bibitem [{\citenamefont {Angenent}\ and\ \citenamefont {Gurtin}(1989)}]{AG89}%
  \BibitemOpen
  \bibfield  {author} {\bibinfo {author} {\bibfnamefont {S.}~\bibnamefont
  {Angenent}}\ and\ \bibinfo {author} {\bibfnamefont {M.~E.}\ \bibnamefont
  {Gurtin}},\ }\href@noop {} {\bibfield  {journal} {\bibinfo  {journal} {Arch.
  Rat. Mech. Anal.}\ }\textbf {\bibinfo {volume} {108}},\ \bibinfo {pages}
  {323} (\bibinfo {year} {1989})}\BibitemShut {NoStop}%
\bibitem [{\citenamefont {Taylor}\ and\ \citenamefont {Cahn}(1994)}]{TC94}%
  \BibitemOpen
  \bibfield  {author} {\bibinfo {author} {\bibfnamefont {J.~E.}\ \bibnamefont
  {Taylor}}\ and\ \bibinfo {author} {\bibfnamefont {J.~W.}\ \bibnamefont
  {Cahn}},\ }\href@noop {} {\bibfield  {journal} {\bibinfo  {journal} {J. Stat.
  Phys.}\ }\textbf {\bibinfo {volume} {77}},\ \bibinfo {pages} {183} (\bibinfo
  {year} {1994})}\BibitemShut {NoStop}%
\bibitem [{\citenamefont {Winning}\ \emph {et~al.}(2001)\citenamefont
  {Winning}, \citenamefont {Gottstein},\ and\ \citenamefont
  {Shvindlerman}}]{Winningetal_AM_2001}%
  \BibitemOpen
  \bibfield  {author} {\bibinfo {author} {\bibfnamefont {M.}~\bibnamefont
  {Winning}}, \bibinfo {author} {\bibfnamefont {G.}~\bibnamefont {Gottstein}},
  \ and\ \bibinfo {author} {\bibfnamefont {L.}~\bibnamefont {Shvindlerman}},\
  }\href@noop {} {\bibfield  {journal} {\bibinfo  {journal} {Acta Mater.}\
  }\textbf {\bibinfo {volume} {49}},\ \bibinfo {pages} {211} (\bibinfo {year}
  {2001})}\BibitemShut {NoStop}%
\bibitem [{\citenamefont {Holm}\ and\ \citenamefont
  {Foiles}(2010)}]{Holmetal_Science_2010}%
  \BibitemOpen
  \bibfield  {author} {\bibinfo {author} {\bibfnamefont {E.}~\bibnamefont
  {Holm}}\ and\ \bibinfo {author} {\bibfnamefont {S.}~\bibnamefont {Foiles}},\
  }\href@noop {} {\bibfield  {journal} {\bibinfo  {journal} {Science}\ }\textbf
  {\bibinfo {volume} {328}},\ \bibinfo {pages} {1138} (\bibinfo {year}
  {2010})}\BibitemShut {NoStop}%
\bibitem [{\citenamefont {Srinivasan}\ \emph {et~al.}(1999)\citenamefont
  {Srinivasan}, \citenamefont {Cahn}, \citenamefont {Jonsson},\ and\
  \citenamefont {Kalonji}}]{Srinivasanetal_AM_1999}%
  \BibitemOpen
  \bibfield  {author} {\bibinfo {author} {\bibfnamefont {S.}~\bibnamefont
  {Srinivasan}}, \bibinfo {author} {\bibfnamefont {J.}~\bibnamefont {Cahn}},
  \bibinfo {author} {\bibfnamefont {H.}~\bibnamefont {Jonsson}}, \ and\
  \bibinfo {author} {\bibfnamefont {G.}~\bibnamefont {Kalonji}},\ }\href@noop
  {} {\bibfield  {journal} {\bibinfo  {journal} {Acta Mater.}\ }\textbf
  {\bibinfo {volume} {47}},\ \bibinfo {pages} {2821} (\bibinfo {year}
  {1999})}\BibitemShut {NoStop}%
\bibitem [{\citenamefont {Upmanyu}\ \emph {et~al.}(2002)\citenamefont
  {Upmanyu}, \citenamefont {Srolovitz}, \citenamefont {Shvindlerman},\ and\
  \citenamefont {Gottstein}}]{Upmanyuetal_AM_2002}%
  \BibitemOpen
  \bibfield  {author} {\bibinfo {author} {\bibfnamefont {M.}~\bibnamefont
  {Upmanyu}}, \bibinfo {author} {\bibfnamefont {D.}~\bibnamefont {Srolovitz}},
  \bibinfo {author} {\bibfnamefont {L.}~\bibnamefont {Shvindlerman}}, \ and\
  \bibinfo {author} {\bibfnamefont {G.}~\bibnamefont {Gottstein}},\ }\href@noop
  {} {\bibfield  {journal} {\bibinfo  {journal} {Acta Mater.}\ }\textbf
  {\bibinfo {volume} {50}},\ \bibinfo {pages} {1405} (\bibinfo {year}
  {2002})}\BibitemShut {NoStop}%
\bibitem [{\citenamefont {Cahn}\ and\ \citenamefont
  {Taylor}(2004)}]{Cahnetal_AM_2004}%
  \BibitemOpen
  \bibfield  {author} {\bibinfo {author} {\bibfnamefont {J.}~\bibnamefont
  {Cahn}}\ and\ \bibinfo {author} {\bibfnamefont {J.}~\bibnamefont {Taylor}},\
  }\href@noop {} {\bibfield  {journal} {\bibinfo  {journal} {Acta Mater.}\
  }\textbf {\bibinfo {volume} {52}},\ \bibinfo {pages} {4887} (\bibinfo {year}
  {2004})}\BibitemShut {NoStop}%
\bibitem [{\citenamefont {Shan}\ \emph {et~al.}(2004)\citenamefont {Shan},
  \citenamefont {Stach}, \citenamefont {Wiezorek}, \citenamefont {Knapp},
  \citenamefont {Follstaadt},\ and\ \citenamefont
  {Mao}}]{Shanetal_Science_2004}%
  \BibitemOpen
  \bibfield  {author} {\bibinfo {author} {\bibfnamefont {Z.}~\bibnamefont
  {Shan}}, \bibinfo {author} {\bibfnamefont {E.}~\bibnamefont {Stach}},
  \bibinfo {author} {\bibfnamefont {J.}~\bibnamefont {Wiezorek}}, \bibinfo
  {author} {\bibfnamefont {J.}~\bibnamefont {Knapp}}, \bibinfo {author}
  {\bibfnamefont {D.}~\bibnamefont {Follstaadt}}, \ and\ \bibinfo {author}
  {\bibfnamefont {S.}~\bibnamefont {Mao}},\ }\href@noop {} {\bibfield
  {journal} {\bibinfo  {journal} {Science}\ }\textbf {\bibinfo {volume}
  {305}},\ \bibinfo {pages} {654} (\bibinfo {year} {2004})}\BibitemShut
  {NoStop}%
\bibitem [{\citenamefont {Upmanyu}\ \emph {et~al.}(2006)\citenamefont
  {Upmanyu}, \citenamefont {Srolovitz}, \citenamefont {Lobkovsky},
  \citenamefont {Warren},\ and\ \citenamefont {Carter}}]{Upmanyuetal_AM_2006}%
  \BibitemOpen
  \bibfield  {author} {\bibinfo {author} {\bibfnamefont {M.}~\bibnamefont
  {Upmanyu}}, \bibinfo {author} {\bibfnamefont {D.}~\bibnamefont {Srolovitz}},
  \bibinfo {author} {\bibfnamefont {A.}~\bibnamefont {Lobkovsky}}, \bibinfo
  {author} {\bibfnamefont {J.}~\bibnamefont {Warren}}, \ and\ \bibinfo {author}
  {\bibfnamefont {W.}~\bibnamefont {Carter}},\ }\href@noop {} {\bibfield
  {journal} {\bibinfo  {journal} {Acta Mater.}\ }\textbf {\bibinfo {volume}
  {54}},\ \bibinfo {pages} {1707} (\bibinfo {year} {2006})}\BibitemShut
  {NoStop}%
\bibitem [{\citenamefont {Trautt}\ and\ \citenamefont
  {Mishin}(2012)}]{Trauttetal_AM_2012}%
  \BibitemOpen
  \bibfield  {author} {\bibinfo {author} {\bibfnamefont {Z.}~\bibnamefont
  {Trautt}}\ and\ \bibinfo {author} {\bibfnamefont {Y.}~\bibnamefont
  {Mishin}},\ }\href@noop {} {\bibfield  {journal} {\bibinfo  {journal} {Acta
  Mater.}\ }\textbf {\bibinfo {volume} {60}},\ \bibinfo {pages} {2407}
  (\bibinfo {year} {2012})}\BibitemShut {NoStop}%
\bibitem [{\citenamefont {Wu}\ and\ \citenamefont
  {Voorhees}(2012)}]{Wuetal_AM_2012}%
  \BibitemOpen
  \bibfield  {author} {\bibinfo {author} {\bibfnamefont {K.-A.}\ \bibnamefont
  {Wu}}\ and\ \bibinfo {author} {\bibfnamefont {P.}~\bibnamefont {Voorhees}},\
  }\href@noop {} {\bibfield  {journal} {\bibinfo  {journal} {Acta Mater.}\
  }\textbf {\bibinfo {volume} {60}},\ \bibinfo {pages} {407} (\bibinfo {year}
  {2012})}\BibitemShut {NoStop}%
\bibitem [{\citenamefont {Heinonen}\ \emph {et~al.}(2014)\citenamefont
  {Heinonen}, \citenamefont {Achim}, \citenamefont {Elder}, \citenamefont
  {Buyukdagli},\ and\ \citenamefont {Ala-Nissila}}]{hein14}%
  \BibitemOpen
  \bibfield  {author} {\bibinfo {author} {\bibfnamefont {V.}~\bibnamefont
  {Heinonen}}, \bibinfo {author} {\bibfnamefont {C.~V.}\ \bibnamefont {Achim}},
  \bibinfo {author} {\bibfnamefont {K.~R.}\ \bibnamefont {Elder}}, \bibinfo
  {author} {\bibfnamefont {S.}~\bibnamefont {Buyukdagli}}, \ and\ \bibinfo
  {author} {\bibfnamefont {T.}~\bibnamefont {Ala-Nissila}},\ }\href@noop {}
  {\bibfield  {journal} {\bibinfo  {journal} {Phys. Rev. E}\ }\textbf {\bibinfo
  {volume} {89}},\ \bibinfo {pages} {032411} (\bibinfo {year}
  {2014})}\BibitemShut {NoStop}%
\bibitem [{\citenamefont {Elder}\ \emph {et~al.}(2002)\citenamefont {Elder},
  \citenamefont {Katakowski}, \citenamefont {Haataja},\ and\ \citenamefont
  {Grant}}]{Elderetal_PRL_2002}%
  \BibitemOpen
  \bibfield  {author} {\bibinfo {author} {\bibfnamefont {K.}~\bibnamefont
  {Elder}}, \bibinfo {author} {\bibfnamefont {M.}~\bibnamefont {Katakowski}},
  \bibinfo {author} {\bibfnamefont {M.}~\bibnamefont {Haataja}}, \ and\
  \bibinfo {author} {\bibfnamefont {M.}~\bibnamefont {Grant}},\ }\href@noop {}
  {\bibfield  {journal} {\bibinfo  {journal} {Phys. Rev. Lett.}\ }\textbf
  {\bibinfo {volume} {88}},\ \bibinfo {pages} {245701} (\bibinfo {year}
  {2002})}\BibitemShut {NoStop}%
\bibitem [{\citenamefont {Elder}\ and\ \citenamefont
  {Grant}(2004)}]{Elderetal_PRE_2004}%
  \BibitemOpen
  \bibfield  {author} {\bibinfo {author} {\bibfnamefont {K.}~\bibnamefont
  {Elder}}\ and\ \bibinfo {author} {\bibfnamefont {M.}~\bibnamefont {Grant}},\
  }\href@noop {} {\bibfield  {journal} {\bibinfo  {journal} {Phys. Rev. E.}\
  }\textbf {\bibinfo {volume} {70}},\ \bibinfo {pages} {051605} (\bibinfo
  {year} {2004})}\BibitemShut {NoStop}%
\bibitem [{\citenamefont {Elder}\ \emph {et~al.}(2007)\citenamefont {Elder},
  \citenamefont {Provatas}, \citenamefont {Berry}, \citenamefont {Stefanovic},\
  and\ \citenamefont {Grant}}]{Elderetal_PRE_2007}%
  \BibitemOpen
  \bibfield  {author} {\bibinfo {author} {\bibfnamefont {K.}~\bibnamefont
  {Elder}}, \bibinfo {author} {\bibfnamefont {N.}~\bibnamefont {Provatas}},
  \bibinfo {author} {\bibfnamefont {J.}~\bibnamefont {Berry}}, \bibinfo
  {author} {\bibfnamefont {P.}~\bibnamefont {Stefanovic}}, \ and\ \bibinfo
  {author} {\bibfnamefont {M.}~\bibnamefont {Grant}},\ }\href@noop {}
  {\bibfield  {journal} {\bibinfo  {journal} {Phys. Rev. E}\ }\textbf {\bibinfo
  {volume} {75}},\ \bibinfo {pages} {064107} (\bibinfo {year}
  {2007})}\BibitemShut {NoStop}%
\bibitem [{\citenamefont {van Teeffelen}\ \emph {et~al.}(2009)\citenamefont
  {van Teeffelen}, \citenamefont {Backofen}, \citenamefont {Voigt},\ and\
  \citenamefont {Loewen}}]{Teeffelenetal_PRE_2009}%
  \BibitemOpen
  \bibfield  {author} {\bibinfo {author} {\bibfnamefont {S.}~\bibnamefont {van
  Teeffelen}}, \bibinfo {author} {\bibfnamefont {R.}~\bibnamefont {Backofen}},
  \bibinfo {author} {\bibfnamefont {A.}~\bibnamefont {Voigt}}, \ and\ \bibinfo
  {author} {\bibfnamefont {H.}~\bibnamefont {Loewen}},\ }\href@noop {}
  {\bibfield  {journal} {\bibinfo  {journal} {Phys. Rev. E}\ }\textbf {\bibinfo
  {volume} {79}},\ \bibinfo {pages} {051404} (\bibinfo {year}
  {2009})}\BibitemShut {NoStop}%
\bibitem [{\citenamefont {Backofen}\ \emph {et~al.}(2014)\citenamefont
  {Backofen}, \citenamefont {Barmak}, \citenamefont {Elder},\ and\
  \citenamefont {Voigt}}]{BBV14}%
  \BibitemOpen
  \bibfield  {author} {\bibinfo {author} {\bibfnamefont {R.}~\bibnamefont
  {Backofen}}, \bibinfo {author} {\bibfnamefont {K.}~\bibnamefont {Barmak}},
  \bibinfo {author} {\bibfnamefont {K.~R.}\ \bibnamefont {Elder}}, \ and\
  \bibinfo {author} {\bibfnamefont {A.}~\bibnamefont {Voigt}},\ }\href@noop {}
  {\bibfield  {journal} {\bibinfo  {journal} {Acta Mater.}\ }\textbf {\bibinfo
  {volume} {64}},\ \bibinfo {pages} {72} (\bibinfo {year} {2014})}\BibitemShut
  {NoStop}%
\bibitem [{\citenamefont {Emmerich}\ \emph {et~al.}(2012)\citenamefont
  {Emmerich}, \citenamefont {Lowen}, \citenamefont {Wittkowski}, \citenamefont
  {Gruhn}, \citenamefont {Tóth}, \citenamefont {Tegze},\ and\ \citenamefont
  {Granasy}}]{Emmerichetal_AP_2012}%
  \BibitemOpen
  \bibfield  {author} {\bibinfo {author} {\bibfnamefont {H.}~\bibnamefont
  {Emmerich}}, \bibinfo {author} {\bibfnamefont {H.}~\bibnamefont {Lowen}},
  \bibinfo {author} {\bibfnamefont {R.}~\bibnamefont {Wittkowski}}, \bibinfo
  {author} {\bibfnamefont {T.}~\bibnamefont {Gruhn}}, \bibinfo {author}
  {\bibfnamefont {G.~I.}\ \bibnamefont {Tóth}}, \bibinfo {author}
  {\bibfnamefont {G.}~\bibnamefont {Tegze}}, \ and\ \bibinfo {author}
  {\bibfnamefont {L.}~\bibnamefont {Granasy}},\ }\href@noop {} {\bibfield
  {journal} {\bibinfo  {journal} {Adv. Phys.}\ }\textbf {\bibinfo {volume}
  {61}},\ \bibinfo {pages} {665} (\bibinfo {year} {2012})}\BibitemShut
  {NoStop}%
\bibitem [{\citenamefont {Faghihi}\ \emph {et~al.}(2013)\citenamefont
  {Faghihi}, \citenamefont {Provatas}, \citenamefont {Elder}, \citenamefont
  {Grant},\ and\ \citenamefont {Karttunen}}]{FPK13}%
  \BibitemOpen
  \bibfield  {author} {\bibinfo {author} {\bibfnamefont {N.}~\bibnamefont
  {Faghihi}}, \bibinfo {author} {\bibfnamefont {N.}~\bibnamefont {Provatas}},
  \bibinfo {author} {\bibfnamefont {K.~R.}\ \bibnamefont {Elder}}, \bibinfo
  {author} {\bibfnamefont {M.}~\bibnamefont {Grant}}, \ and\ \bibinfo {author}
  {\bibfnamefont {M.}~\bibnamefont {Karttunen}},\ }\href@noop {} {\bibfield
  {journal} {\bibinfo  {journal} {Phys. Rev. E}\ }\textbf {\bibinfo {volume}
  {88}},\ \bibinfo {pages} {032407} (\bibinfo {year} {2013})}\BibitemShut
  {NoStop}%
\bibitem [{\citenamefont {Seymour}\ \emph {et~al.}(2015)\citenamefont
  {Seymour}, \citenamefont {Sanches}, \citenamefont {Elder},\ and\
  \citenamefont {Provatas}}]{SSP15}%
  \BibitemOpen
  \bibfield  {author} {\bibinfo {author} {\bibfnamefont {M.}~\bibnamefont
  {Seymour}}, \bibinfo {author} {\bibfnamefont {F.}~\bibnamefont {Sanches}},
  \bibinfo {author} {\bibfnamefont {K.}~\bibnamefont {Elder}}, \ and\ \bibinfo
  {author} {\bibfnamefont {N.}~\bibnamefont {Provatas}},\ }\href@noop {}
  {\bibfield  {journal} {\bibinfo  {journal} {Phys. Rev. B}\ }\textbf {\bibinfo
  {volume} {92}},\ \bibinfo {pages} {184109} (\bibinfo {year}
  {2015})}\BibitemShut {NoStop}%
\bibitem [{\citenamefont {Greenwood}\ \emph {et~al.}(2010)\citenamefont
  {Greenwood}, \citenamefont {Provatas},\ and\ \citenamefont
  {Rottler}}]{GPR10}%
  \BibitemOpen
  \bibfield  {author} {\bibinfo {author} {\bibfnamefont {M.}~\bibnamefont
  {Greenwood}}, \bibinfo {author} {\bibfnamefont {N.}~\bibnamefont {Provatas}},
  \ and\ \bibinfo {author} {\bibfnamefont {J.}~\bibnamefont {Rottler}},\
  }\href@noop {} {\bibfield  {journal} {\bibinfo  {journal} {Phys. Rev. Lett.}\
  }\textbf {\bibinfo {volume} {105}},\ \bibinfo {pages} {045702} (\bibinfo
  {year} {2010})}\BibitemShut {NoStop}%
\bibitem [{\citenamefont {Ofori-Opoku}\ \emph {et~al.}(2013)\citenamefont
  {Ofori-Opoku}, \citenamefont {Stolle}, \citenamefont {Huang},\ and\
  \citenamefont {Provatas}}]{OSP13}%
  \BibitemOpen
  \bibfield  {author} {\bibinfo {author} {\bibfnamefont {N.}~\bibnamefont
  {Ofori-Opoku}}, \bibinfo {author} {\bibfnamefont {J.}~\bibnamefont {Stolle}},
  \bibinfo {author} {\bibfnamefont {Z.-F.}\ \bibnamefont {Huang}}, \ and\
  \bibinfo {author} {\bibfnamefont {N.}~\bibnamefont {Provatas}},\ }\href@noop
  {} {\bibfield  {journal} {\bibinfo  {journal} {Phys. Rev. B}\ }\textbf
  {\bibinfo {volume} {88}},\ \bibinfo {pages} {104106} (\bibinfo {year}
  {2013})}\BibitemShut {NoStop}%
\bibitem [{\citenamefont {Molodov}\ and\ \citenamefont
  {Konijnenberg}(2006)}]{Molodovetal_SM_2006}%
  \BibitemOpen
  \bibfield  {author} {\bibinfo {author} {\bibfnamefont {D.~A.}\ \bibnamefont
  {Molodov}}\ and\ \bibinfo {author} {\bibfnamefont {P.~J.}\ \bibnamefont
  {Konijnenberg}},\ }\href@noop {} {\bibfield  {journal} {\bibinfo  {journal}
  {Scr. Mater.}\ }\textbf {\bibinfo {volume} {54}},\ \bibinfo {pages} {977}
  (\bibinfo {year} {2006})}\BibitemShut {NoStop}%
\bibitem [{\citenamefont {Barrales-Mora}\ \emph {et~al.}(2007)\citenamefont
  {Barrales-Mora}, \citenamefont {Mohles}, \citenamefont {Konijnenberg},\ and\
  \citenamefont {Molodov}}]{Barrales-Moraetal_CMS_2007}%
  \BibitemOpen
  \bibfield  {author} {\bibinfo {author} {\bibfnamefont {L.}~\bibnamefont
  {Barrales-Mora}}, \bibinfo {author} {\bibfnamefont {V.}~\bibnamefont
  {Mohles}}, \bibinfo {author} {\bibfnamefont {P.~J.}\ \bibnamefont
  {Konijnenberg}}, \ and\ \bibinfo {author} {\bibfnamefont {D.~A.}\
  \bibnamefont {Molodov}},\ }\href@noop {} {\bibfield  {journal} {\bibinfo
  {journal} {Comput. Mater. Sci.}\ }\textbf {\bibinfo {volume} {39}},\ \bibinfo
  {pages} {160} (\bibinfo {year} {2007})}\BibitemShut {NoStop}%
\bibitem [{\citenamefont {Rivoirard}(2013)}]{Rivoirard_JOM_2013}%
  \BibitemOpen
  \bibfield  {author} {\bibinfo {author} {\bibfnamefont {S.}~\bibnamefont
  {Rivoirard}},\ }\href@noop {} {\bibfield  {journal} {\bibinfo  {journal}
  {JOM}\ }\textbf {\bibinfo {volume} {65}},\ \bibinfo {pages} {901} (\bibinfo
  {year} {2013})}\BibitemShut {NoStop}%
\bibitem [{\citenamefont {Sekerka}(2005)}]{Sek05}%
  \BibitemOpen
  \bibfield  {author} {\bibinfo {author} {\bibfnamefont {R.~F.}\ \bibnamefont
  {Sekerka}},\ }\href {\doibase 10.1002/crat.200410342} {\bibfield  {journal}
  {\bibinfo  {journal} {Cryst. Res. Technol.}\ }\textbf {\bibinfo {volume}
  {40}},\ \bibinfo {pages} {291} (\bibinfo {year} {2005})}\BibitemShut
  {NoStop}%
\bibitem [{\citenamefont {Han}\ \emph {et~al.}(2018)\citenamefont {Han},
  \citenamefont {Thomas},\ and\ \citenamefont {Srolovitz}}]{Hanetal_PMS_2018}%
  \BibitemOpen
  \bibfield  {author} {\bibinfo {author} {\bibfnamefont {J.}~\bibnamefont
  {Han}}, \bibinfo {author} {\bibfnamefont {S.}~\bibnamefont {Thomas}}, \ and\
  \bibinfo {author} {\bibfnamefont {D.}~\bibnamefont {Srolovitz}},\ }\href@noop
  {} {\bibfield  {journal} {\bibinfo  {journal} {Prog. Mater. Sci,}\ }\textbf
  {\bibinfo {volume} {98}},\ \bibinfo {pages} {386} (\bibinfo {year}
  {2018})}\BibitemShut {NoStop}%
\bibitem [{\citenamefont {K{\"{o}}hler}\ \emph {et~al.}(2016)\citenamefont
  {K{\"{o}}hler}, \citenamefont {Backofen},\ and\ \citenamefont
  {Voigt}}]{Koehleretal_PRL_2017}%
  \BibitemOpen
  \bibfield  {author} {\bibinfo {author} {\bibfnamefont {C.}~\bibnamefont
  {K{\"{o}}hler}}, \bibinfo {author} {\bibfnamefont {R.}~\bibnamefont
  {Backofen}}, \ and\ \bibinfo {author} {\bibfnamefont {A.}~\bibnamefont
  {Voigt}},\ }\href {http://link.aps.org/doi/10.1103/PhysRevLett.116.135502}
  {\bibfield  {journal} {\bibinfo  {journal} {Phys. Rev. Lett.}\ }\textbf
  {\bibinfo {volume} {116}},\ \bibinfo {pages} {1} (\bibinfo {year}
  {2016})}\BibitemShut {NoStop}%
\bibitem [{\citenamefont {G\"unster}\ \emph {et~al.}(2010)\citenamefont
  {G\"unster}, \citenamefont {Molodov},\ and\ \citenamefont
  {Gottstein}}]{Guensteretal_SM_2010}%
  \BibitemOpen
  \bibfield  {author} {\bibinfo {author} {\bibfnamefont {C.}~\bibnamefont
  {G\"unster}}, \bibinfo {author} {\bibfnamefont {D.~A.}\ \bibnamefont
  {Molodov}}, \ and\ \bibinfo {author} {\bibfnamefont {G.}~\bibnamefont
  {Gottstein}},\ }\href@noop {} {\bibfield  {journal} {\bibinfo  {journal}
  {Scr. Mater.}\ }\textbf {\bibinfo {volume} {63}},\ \bibinfo {pages} {300}
  (\bibinfo {year} {2010})}\BibitemShut {NoStop}%
\end{thebibliography}%

\newpage

\clearpage


\end{document}